\documentclass[a4paper,showpacs,prl,twocolumn]{revtex4}
\usepackage{amsfonts}
\usepackage{graphicx}
\usepackage{color}
\usepackage{amsmath}
\usepackage{amssymb}
\usepackage{latexsym}
\usepackage{psfrag}

\begin{document}

\title{The quenching of compressible edge states around antidots}
\author{S. Ihnatsenka}
\altaffiliation{Permanent address: Centre of Nanoelectronics,
Department of Microelectronics, Belarusian State University for
Informatics and Radioelectronics, 220013 Minsk, Belarus}
\affiliation{Solid State Electronics, Department of Science and Technology (ITN), Link%
\"{o}ping University, 60174 Norrk\"{o}ping, Sweden}
\date{\today }
\author{I. V. Zozoulenko}
\affiliation{Solid State Electronics, Department of Science and Technology (ITN), Link%
\"{o}ping University, 60174 Norrk\"{o}ping, Sweden}
\date{\today }

\begin{abstract}
We provide a systematic quantitative description of the edge state
structure around a quantum antidot in the integer quantum Hall
regime. The calculations for spinless electrons within the Hartree
approximation reveal that the widely used Chklovskii \textit{et
al.} electrostatic description greatly overestimates the widths of
the compressible strips; the difference between these approaches
diminishes as the size of the antidot increases. By including spin
effects within density functional theory in the local spin-density
approximation, we demonstrate that the exchange interaction can
suppress the formation of compressible strips and lead to a
spatial separation between the spin-up and spin-down states. As
the magnetic field increases, the outermost compressible strip,
related to spin-down states starts to form. However, in striking
contrast to quantum wires, the innermost compressible strip (due
to spin-up states) never develops for antidots.
\end{abstract}

\maketitle

A quantum antidot is a potential hill in a two-dimensional electron gas
(2DEG) usually defined by means of an electrostatic split gate, see Fig. \ref%
{fig:structure}. In a perpendicular magnetic field, electrons are
trapped around the antidot in bound states formed by magnetic
confinement. Experimental studies of magnetotransport in quantum
antidots reported over the last decade reveal a rich
magneto-conductance structure in the
quantum Hall regime \cite{Andy,Goldman_Science,Maasilta,Karakurt,Goldman2005,%
Ford_1994,Mace,Kataoka_1999,Kataoka_2000,Kataoka_2002,Kataoka_2003}.
Some of the observed magneto-conductance features can be
understood within a one-electron picture in terms of
semi-classical and quantum electron dynamics \cite{Andy}. However,
a majority of experiments confirm a central role played by
electron interactions and spin effects in antidot measurements.
This includes, for example, the first direct observation of the
fractionally quantized electron charge \cite{Goldman_Science},
fractional statistics \cite{Goldman2005}, the striking effect of
the frequency doubling of the Aharonov-Bohm (AB) oscillations \cite%
{Ford_1994,Kataoka_2000}, the detection of the Coulomb charging \cite%
{Kataoka_1999}, the observation of the Kondo effect
\cite{Kataoka_2002} and selective spin-injection
\cite{Kataoka_2003}. Interest in antidot structures is also
motivated by their potential for spintronic applications, where
they can be used to inject or detect spin-polarized currents
\cite{APL} or even as quantum gates \cite{Berggren}. The antidots
also provide a system for investigating edge states in general,
because of the detailed information one can obtain from the
dependence of AB peak positions on field and gate voltage.

A detailed microscopic understanding of antidot systems therefore
requires a rigorous theory accounting for both interaction and
spin effects. In contrast to quantum dots and wires which have
been the subject of intense theoretical study (see e.g.
\cite{QDOverview,Ihnatsenka} and references therein), the
energetics of quantum antidots has received, with only a few exceptions \cite%
{Hwang}, practically no attention. In particular, the central
issue concerning the structure of edge states and the formation of
the compressible strips around antidots still remains an open
question. The structure of the antidot edge states represents an
important key to an understanding of various effects such as AB
oscillations \cite{Ford_1994,Kataoka_2000}, Coulomb charging
\cite{Kataoka_1999} and spin selectivity \cite{Kataoka_2003},
and it has been a subject of recent lively discussions \cite%
{Karakurt,PRL_comments}. The main goal of the present paper is to
provide a rigorous theoretical description for the spin-resolved structure of
edge states around a quantum antidot.

%*********************************************************
\begin{figure}[tb]
\includegraphics[scale=1.0]{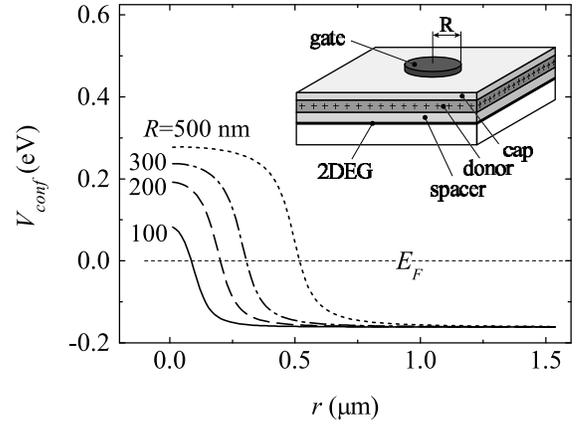} %[tph]
\caption{Calculated electrostatic confinement potentials
$V_{conf}(r)$ for a quantum antidot with different gate radii $R$.
The inset shows a schematical layout of the device defined in GaAs
heterostructure with a cap layer of the width $c=14$ nm, a donor
layer of the width $d=36$
nm, and a spacer layer of the width $s=10$ nm. The donor concentration $%
n_{d}=6\cdot 10^{23}$ m$^{-3}$. The above parameters (corresponding to the
bulk electron density $n_{bulk}\sim 2\cdot 10^{15}\text{m}^{-2}$) are used
for all antidot structures discussed in the paper.}
\label{fig:structure}
\end{figure}
%*********************************************************
We consider an antidot defined within a GaAs heterostructure by a circular gate
of radius $R$ as illustrated in the inset to Fig. \ref{fig:structure}.
We assume that electron motion is confined to the plane parallel to the
heterostructure interface, $\mathbf{r}=(x,y).$ The external electrostatic
confinement, $V_{conf}(r)=V_{Schottky}+V_{d}+V_{g}(r),$ includes the
Schottky barrier $V_{Schottky}=0.8$ eV and the potential due to a layer of
donors of width $d$ situated at a distance $c$ from the surface \cite%
{Martorell}, $V_{d}=-n_{d}d\left( c+d/2\right) e^{2}/\varepsilon
_{0}\varepsilon _{r},$ with $n_{d}$ being the donor concentration and $%
\varepsilon _{r}$ the GaAs dielectric constant. The electrostatic
potential $V_{g}(r)$ due to a circular gate is given by an analytical
expression provided by Davies (Eq. (3.17) in Ref. \onlinecite{Davies}). The
electrostatic confinement $V_{conf}(r)$ for different gate radii $R$ is
shown in Fig. \ref{fig:structure} .

Utilizing the circular symmetry of the structure we introduce cylindrical
coordinates and write down the wave function in the form $\psi
(r)=e^{il\varphi }\phi (r),$ where $l=0,\pm 1,\ldots $ is the orbital
quantum number. The Schr\"{o}dinger equation in a perpendicular magnetic
field $B$ reads
\begin{equation}
-\frac{1}{r}\frac{\partial }{\partial r}r\frac{\partial \phi ^{\sigma }(r)}{%
\partial r}+\left[ \left( \frac{l}{r}+\frac{qr}{2}\right) ^{2}+V^{\sigma }(r)%
\right] \phi ^{\sigma }(r)=E\phi ^{\sigma }(r),  \label{Schrodinger}
\end{equation}%
where the lengths are measured in units of the lattice constant $a$ (used for
numerical discretization), energies in units of $\hbar ^{2}/2m^{\ast }a^{2};$
$q=eBa^{2}/\hbar ,$ and $\sigma =\pm
%TCIMACRO{\U{bd}}%
%BeginExpansion
{\frac12}%
%EndExpansion
$ describes spin-up and spin-down states, $\uparrow ,\downarrow $.
We include electron interactions and spin effects within the
framework of density functional theory (DFT) in the local spin
density approximation (LSDA) \cite{ParrYang}. The choice of
DFT+LSDA for the description of many-electron effects is
motivated, on one hand, by its practical implementation efficiency
within a standard Kohn-Sham formalism \cite{Kohn}, and on the
other hand, by the excellent agreement between the DFT+LSDA and
exact diagonalization \cite{Stephanie} and variational Monte-Carlo
calculations \cite{Rasanen,QDOverview} performed for few-electron
systems. Within the framework of the DFT+LSDA, the total
confinement potential can be
written in the form%
\begin{equation}
V^{\sigma }(r)=V_{conf}(r)+V_{H}(r)+V_{xc}^{\sigma }(r)+g\mu _{B}B\sigma ,
\label{V}
\end{equation}%
where
\begin{equation}
V_{H}(r)=\frac{e^{2}}{4\pi \varepsilon _{0}\varepsilon _{r}}\int d\mathbf{r}%
^{\prime }n(\mathbf{r}^{\prime })\left[ \frac{1}{\left\vert \mathbf{r}-%
\mathbf{r}^{\prime }\right\vert }-\frac{1}{\sqrt{\left( \mathbf{r}-\mathbf{r}%
^{\prime }\right) ^{2}+4b^{2}}}\right]   \label{V_H}
\end{equation}%
is the Hartree potential including the contribution from mirror charges ($b$
is the distance from the 2DEG to the surface), $n(r)=\sum_{\sigma }n^{\sigma
}(r);\;n^{\sigma }(r)=\sum_{i}\left\vert \psi _{i}^{\sigma }(r)\right\vert
^{2}f_{FD}(E-E_{F})$ is the electron density, and $f_{FD}$ is the
Fermi-Dirac distribution. For the exchange and correlation potential $%
V_{xc}^{\sigma }(r)$ we utilize a widely used parameterization from Tanatar
and Cerperly \cite{TC} (see Ref. \onlinecite{Ihnatsenka} for the explicit
expressions for $V_{xc}(r)$). This parameterization is valid for magnetic
fields corresponding to filling factors $\nu >1$, which sets the limit for the
applicability of our results. The last term in Eq. (\ref{V}) accounts for
the Zeeman energy where $\mu _{b}=\frac{e\hbar }{2m_{e}}$ is the Bohr magnetron,
and the bulk $g$ factor of GaAs is $g=-0.44$. We solve Eq. (\ref{Schrodinger}%
) self-consistently expanding the wavefunctions into sin-basis. Because the
antidot represents an open system, we choose the computational
domain sufficiently large to ensure that the electron density in the bulk
(i.e. far away from the antidot) is constant and does not change when we
increase the domain size.

\begin{figure}[tb]
\includegraphics[scale=1.0]{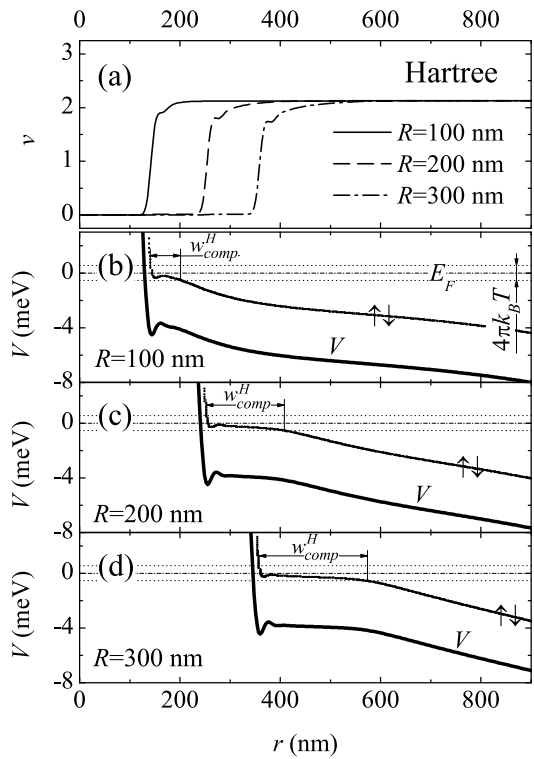} %[tph]
\caption{(a) The electron density profile (local filling factor), $\protect%
\nu (r)=n(r)/n_{B}$ for antidots of different radii $R$ calculated
in the Hartree approximation. (b) The corresponding magnetosubband
structure (i.e. the overage position of the wave functions
$\protect\psi _{i}$). Fat solid lines indicate the total confining
potential $V(r)$, Eq. (\protect\ref{V}). Magnetic field $B=4.2$ T.
Temperature $T=1$ K.} \label{fig:band_Hartree}
\end{figure}

Following our previous analysis of edge state structure in quantum wires
\cite{Ihnatsenka,Ihnatsenka2,Ihnatsenka3}, we start with the
Hartree approximation [disregarding exchange and correlation interactions by
setting $V_{xc}^{\sigma }(r)=0$ in Eq. (\ref{V}) ]. Figure \ref%
{fig:band_Hartree} (a) shows the electron density profiles (the local
filling factors) $\nu (r)=n(r)/n_{B}$ $(n_{B}=eB/h)$ around antidots with
different radii for a representative value of magnetic field $B=4.2$ T. The
corresponding average wave function positions $\psi _{i}$ for
different eigenenergies $E_{i}$ (i.e. the magnetosubbands) are shown in
Figs. \ref{fig:band_Hartree} (b)-(d) illustrating the formation of the
compressible strips around the antidots. [Note that in Fig. \ref%
{fig:band_Hartree} each eigenstate $\psi _{i}$ is represented by a dot.
Because of a large number of the eigenstates, $\sim 10^{4},$ the dots are
merged into solid lines]. The compressible strips are composed of partially
filled electron states that screen the external potential and lead to a
flattening of the subbands in the compressible regions \cite{Chklovskii}.
[Following Refs. \cite{Ando,Ihnatsenka,Ihnatsenka2,Ihnatsenka3} We define
the compressible strips within the window $\left\vert E-E_{F}<2\pi
kT\right\vert $)]. Figure \ref{fig:band_Hartree} also shows the total
confining potential $V(r)$, Eq. (\ref{V}). Note that the calculated spin-up
and spin-down densities and potentials are virtually indistinguishable on
the scale of the figure. Hence, in this magnetic field interval
the effect of the Zeeman term on the subband structure is negligible, so that
we may refer to the Hartree results as being the case of spinless electrons.
%*********************************************************
%*********************************************************
%*********************************************************
\begin{figure}[tb]
\includegraphics[scale=1.0]{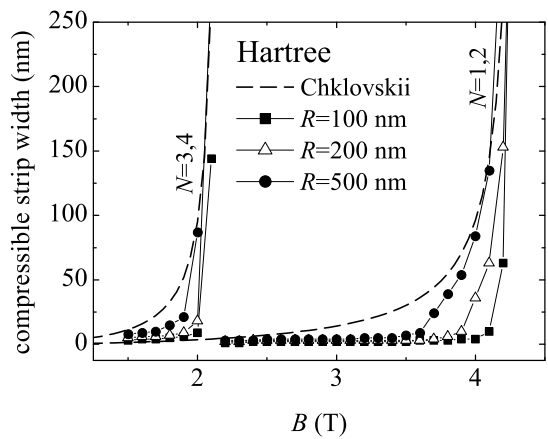} %[tph]
\caption{Width of the compressible strips around quantum antidot
with different radii $R$ as a function of magnetic field
calculated within the Hartree approximation (i.e. for the spinless
electrons) and its comparison to the Chklovskii \textit{et al. }
predictions \protect\cite{Chklovskii}. $N=1,2$ and 3,4 refer to
the subband number corresponding to the two lowest (spin
degenerate) edge states. Temperature $T=1$ K.}
\label{fig:comp_Hartree}
\end{figure}
%*********************************************************

Figure \ref{fig:comp_Hartree} shows the width of the compressible strips for
spinless electrons, $w_{comp}^{H}$, around antidots of different radii. For
a comparison, we also plot Chklovskii \textit{et al. } \cite{Chklovskii}
analytical expressions, $w_{comp}^{Chk},$ giving the length of the
compressible strips at the edge of a semi-infinite 2DEG. [$w_{comp}^{Chk}$
depends on two parameters, the depletion length $l,$ and the filling factor
in the bulk, $\nu _{bulk}=n_{bulk}/n_{B}.$ We extract $l$ from the
calculated self-consistent density distribution by fitting to the dependence
$n(r)=n_{bulk}\left( \frac{r-l}{r+l}\right) ^{1/2}$where $n_{bulk}$ is the
electron density far away from the antidot \cite{Chklovskii,Ando,Ihnatsenka2}
]. It has been demonstrated that for the case of quantum wires the width of
the compressible strips for spinless electrons calculated in the Hartree
approach, $w_{comp}^{H},$ is in very good agreement with the Chklovskii
\textit{et al. }predictions \cite{Chklovskii} for $w_{comp}^{Chk}$ \cite%
{Ando,Ihnatsenka2}\textit{.} This is obviously not the case for
the antidot structures where the Chklovskii \textit{et al.
}predictions \cite{Chklovskii} greatly overestimates the
compressible strip width. For example, for an antidot with the
radius $R=200$ nm, the innermost compressible strip (i.e.
corresponding to the edge state closest to the antidot) starts to
form at $B\approx 3.9$ T, whereas for the 2DEG edge this
strip starts to already form at $B\approx 2.6$ T, see Fig. \ref%
{fig:comp_Hartree}. The difference between $w_{comp}^{H}$ and
$w_{comp}^{Chk} $ is most pronounced for small antidot radii $R$
and decreases as $R$ increases (note that the limit $R\rightarrow
\infty $ effectively corresponds to the case of a straight
boundary, i.e. the semi-infinite 2DEG). The difference between
$w_{comp}^{H}$ and $w_{comp}^{Chk}$ can be understood as follows.
For the case of a semi-infinite gate an electron in the vicinity
of the edge of the 2DEG experiences the Hartree potential
originated from electrons in the semi-infinite region not covered
by the gate. However, for the case of the antidot the Hartree
potential is stronger as it includes an additional contribution
from the electrons surrounding the antidot (that are otherwise
depleted for the case of the semi-infinite 2DEG). This additional
contribution effectively repels the electrons from the boundary
towards the bulk of the 2DEG. This leads to less effective
screening and thus to steeper potential preventing the formation
of compressible strips.

%*********************************************************
\begin{figure*}[tb]
\includegraphics[scale=0.9]{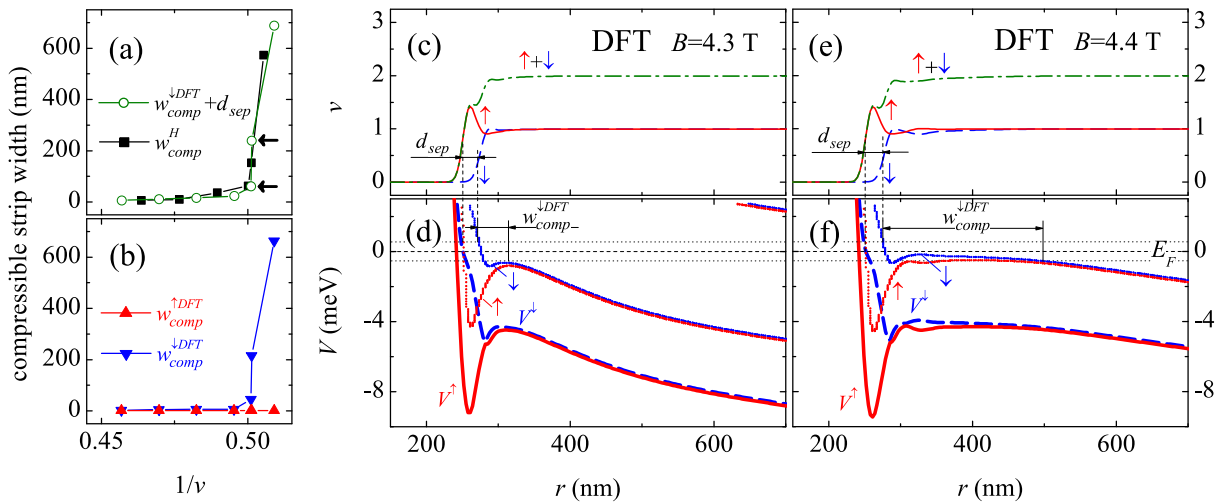} %[tph]
\caption{(Color online) (a) Width of the compressible strips for
spinless electrons in the Hartree approximation, $w_{comp}^{H}$,
as compared to $d_{sep}+w_{comp}^{\downarrow DFT}$ calculated
within DFT. (b) With of the compressible strips for spin-up and
spin-down states $w_{comp}^{\uparrow DFT}, w_{comp}^{\downarrow
DFT}$ calculated within DFT.
 (c), (e) The electron density profiles (local filling factor), $%
\protect\nu^\protect\sigma (r)=n^\protect\sigma (y)/n_{B}$ for
different magnetic fields (indicated by arrows in (a)) $B=4.3$ T
and $4.4$ T, and, (d), (f) the corresponding magnetosubband
structure (i.e. the overage position of the wave functions
$\protect\psi^\protect\sigma _{i}$). Fat solid lines indicate the
total confining potential $V^\protect\sigma(r)$ , Eq.
(\protect\ref{V}). $T=1$ K. } \label{fig:DFT}
\end{figure*}
%*********************************************************

Let us now analyze the spin-resolved edge state structure within the DFT
approximation. Figures \ref{fig:DFT} (c)-(f) show the electron density
profiles (the local filling factors) $\nu ^{\sigma }(r)=n^{\sigma }(r)/n_{B}$
and the average wave function position $\psi _{i}^{\sigma }$ for
different energies $E_{i}^{\sigma }$ (the magnetosubband structure) around the
antidots with radius $R=200$ nm for different representative magnetic fields
$B=4.3$ T and $B=4.4$ T. Within the Hartree approximation the subbands are
virtually degenerate since the Zeeman splitting is very small in the
magnetic field interval under investigation (see Fig. \ref{fig:band_Hartree}%
). In contrast, the exchange interaction included within the DFT
approximation causes the separation of the subbands for spin-up and
spin-down electrons. Indeed, the exchange potential for spin-up electrons
depends on the density of spin-down electrons and vice versa \cite%
{ParrYang,TC,Ihnatsenka}. In compressible regions the subbands
are only partially filled (because $f_{FD}<1$ in the the window
$|E-E_{F}|\lesssim 2\pi kT$), and, therefore, the population of
spin-up and spin-down subbands may be different. In the DFT
calculation, this population difference (triggered by Zeeman
splitting) is strongly enhanced by the exchange interaction
leading to different effective potentials for spin-up and
spin-down electrons and eventually to a subband spin splitting.
As a result, the compressible region present in the Hartree
approximation is suppressed and the spin-up and spin-down states
become spatially separated by the distance $d_{sep}\approx
w_{comp}^{H}$, see Figs. \ref{fig:DFT} (a)-(d). On further
increasing the magnetic field the compressible strip starts to
form for the outer (spin-down) state such that
$d_{sep}+w_{comp}^{\downarrow DFT}\approx w_{comp}^{H},$ see Figs. \ref%
{fig:DFT} (a)--(b),(e)-(f) ($w_{comp}^{\sigma DFT}$ is the width of the
compressible strip for the spin state $\sigma $ calculated in the DFT
approximation).

Far away from the antidot the subbands remain degenerate since they are
situated below the Fermi energy $E\lesssim E_{F}-2\pi kT$ and are thus fully
occupied ($f_{FD}=1$). As a result, the corresponding spin-up and spin-down
densities are the same, hence the exchange and correlation potentials for
the spin-up and spin-down electrons are equal, $V_{xc}^{\uparrow
}(r)=V_{xc}^{\downarrow }(r)$.

A similar scenario for subband spin splitting also holds for
quantum wires \cite{Ihnatsenka2}. However, an important and
interesting distinction is that for quantum wires, as the magnetic
field is increased, compressible strips form first for spin-up and
then for spin-down states. In contrast, for the case of the
antidot only the compressible strip for the spin-down state forms,
whereas the compressible strip for the spin-up states (situated
close to the antidot) never develops, see Fig. \ref{fig:DFT}(b)
[This conclusion holds for all antidots sizes studied in this
paper, see Fig. \ref{fig:structure}]. Just as for the case of
spinless electrons discussed above, we attribute this difference
to less effective screening for the antidot structure leading to a
rather steep potential near the antidot boundary that prevents
formation of the compressible strip for the innermost (spin-up)
state. Note that the absence of the compressible strip for the
inner (spin-up) state is consistent with the interpretation of the
unexpected doubling of the frequency of the Aharonov-Bohm
oscillations explained in Ref. \cite{Kataoka_2000} in terms of a
charging of the outermost compressible regions around the antidots
by electrons of the same spin. Our calculations indicate that the
charging of the innermost (spin-up states) is unlikely since the
spin-up states do not form the compressible regions. [We note
however, that we are not yet in a position to comment
on whether charging of the compressible strips proposed in Ref. \cite%
{Kataoka_2000} really does takes place. The answer to this question can be
obtained from self-consistent \textit{transport }calculations similar to
those reported in e.g. Ref. \cite{open_dot}. Such calculations are currently
in progress.]

Finally we stress that all the results and conclusions presented
in this paper for a temperature $T=1$ K remain valid for lower
temperatures, since calculations performed for $T=0.2$ K reveal
that the width of the compressible strips remain practically
unchanged.

It is important to note that the effect of reduced screening that
strongly affects the antidot edge structure as discussed above
might also be imperative for the case of quantum dot, where the
spin selectivity in the edge state regime might strongly depend on
the gate layout \cite{Ciorda}. The implications of this effect for
quantum dot geometries remains to be identified.

To conclude, we find that for spinless electrons edge states
around an antidot start to form for magnetic fields significantly
higher than those
predicted by Chklovskii \textit{et al.} electrostatic description \cite%
{Chklovskii}. By including spin effects within spin density
functional theory we show that the exchange interaction leads to
qualitatively novel features in antidot edge state structure, such
as the suppression of compressible strips for lower fields,
spatial separation between spin-up and spin down states, and to
the total absence of the innermost compressible strip due to
spin-up states (as opposed to the outermost compressible strip due
to spin-down states that forms at higher fields).

S. I. acknowledges financial support from the Swedish Institute.
Discussions with C. J. B. Ford, M. Kataoka and A. S. Sachrajda are
greatly appreciated. We thankful to C. J. B. Ford and A. S.
Sachrajda for critical reading of the manuscript and valuable
suggestions.

\end{document}